\documentclass[useAMS,usenatbib]{mn2e}

\usepackage{graphicx}

\usepackage{tabularx}

\title[Multiple Populations in Globular Clusters and the Origin of the Oosterhoff Period Groups]{Multiple Populations in Globular Clusters and the Origin of the Oosterhoff Period Groups}
\author[Sohee Jang, Young-Wook Lee, Seok-Joo Joo, and Chongsam Na]{Sohee Jang, Young-Wook
Lee\thanks{E-mail: ywlee@csa.yonsei.ac.kr}, Seok-Joo Joo, and Chongsam Na\\
Center for Galaxy Evolution Research and Department of Astronomy, 
	Yonsei University, Seoul 120-749, Korea}

\begin{document}

\date{Accepted 2014 february 00. Received 2014 february 00; in original form 2014 february 00}

\pagerange{\pageref{firstpage}--\pageref{lastpage}} \pubyear{2014}

\maketitle

\label{firstpage}
\begin{abstract}
The presence of multiple populations is now well-established in most globular clusters in the Milky Way. In light of this progress, here we suggest a new model explaining the origin of the Sandage period-shift and the difference in mean period of type ab RR Lyrae variables between the two Oosterhoff groups. In our models, the instability strip in the metal-poor group II clusters, such as M15, is populated by second generation stars (G2) with enhanced helium and CNO abundances, while the RR Lyraes in the relatively metal-rich group I clusters like M3 are mostly produced by first generation stars (G1) without these enhancements. This population shift within the instability strip with metallicity can create the observed period-shift between the two groups, since both helium and CNO abundances play a role in increasing the period of RR Lyrae variables. The presence of more metal-rich clusters having Oosterhoff-intermediate characteristics, such as NGC 1851, as well as of most metal-rich clusters having RR Lyraes with longest periods (group III) can also be reproduced, as more helium-rich third and later generations of stars (G3) penetrate into the instability strip with further increase in metallicity. Therefore, although there are systems where the suggested population shift cannot be a viable explanation, for the most general cases, our models predict that the RR Lyraes are produced mostly by G1, G2, and G3, respectively, for the Oosterhoff groups I, II, and III.
\end{abstract}

%%##############################################################################################################################

%%##############################################################################################################################
\section{Introduction}

One of the long-standing problems in modern astronomy is the curious division of globular clusters (GCs) into two groups, according to the mean period ($\left<P_{\rm ab}\right>$) of type ab RR Lyrae variables \citep{Oos39}. Understanding this phenomenon, ``the Oosterhoff dichotomy'', is intimately related to the population II distance scale and the formation of the Milky Way halo 
\citep[][and references therein]{San81,Lee90,Yoo02,Cat09}.

\citet{van73} first suggested ``hysteresis mechanism", which explains the dichotomy as a difference in mean temperature between the type ab RR Lyraes in two groups. That this cannot be the whole explanation for the dichotomy became clear when \citet{San81} found a period-shift at given temperature between the RR Lyraes in two GCs representing each of the Oosterhoff groups, M15 and M3. Sandage suggested that this shift is due to the luminosity difference, which, however, required that the RR Lyraes in M15 are abnormally enhanced in helium abundance. \citet[][hereafter LDZ I]{Lee90}, on the other hand, found that RR Lyraes evolved away from the zero-age horizontal-branch (ZAHB) can explain the observed period-shift when the HB type \citep[][hereafter LDZ II]{Lee94} is sufficiently blue. In the case of M15, which has a blue tail (or extreme blue HB; EBHB) in addition to normal blue HB \citep{Buo85}, it was not clear though whether this evolution effect alone can reproduce the period-shift. More recent colour-magnitude diagrams (CMDs) for other metal-poor Oosterhoff group II GCs, NGC~4590, 5053, and 5466 \citep{Wal94,Nem04,Cor99}, also show that the HB types for these GCs are too red (HB~Type $\approx$ 0.5) to have enough evolution effect. This suggests that a significant fraction of RR Lyraes in these GCs, including M15, are probably near the ZAHB. Therefore, the complete understanding of the difference between the two Oosterhoff groups still requires further investigation.

Recent discovery of multiple populations in GCs is throwing new light on this problem. Even in ``normal'' GCs without signs of supernovae enrichments, observations and population models suggest the presence of two or more subpopulations differing in helium and light elements abundances, including CNO \citep[][and references therein]{Gra12a}. It was suggested to be due to the chemical pollution and enrichment by intermediate-mass asymptotic giant branch (AGB) stars, fast-rotating massive stars, and/or rotating AGB stars \citep{Ven09,Dec07,Dec09}. Since the colour of the HB is sensitively affected by age, helium and CNO abundances (see LDZ II), each subpopulation in a GC would be placed in a different colour regime on the HB. Similarly, this would affect the period of RR Lyraes as the variation in chemical composition would change the luminosity and mass of a HB star within the instability strip, by which the period is determined when temperature is fixed \citep{van73}. The purpose of this Letter is to suggest that, in the multiple populations paradigm, the difference in period between the two Oosterhoff groups can be reproduced as the instability strip is progressively occupied by different subpopulations with increasing metallicity.
%%##############################################################################################################################

%%##############################################################################################################################
\section{Population shift within the instability strip}

\begin{figure}
\centering
\vspace{-10pt}
\includegraphics[width=0.53\textwidth]{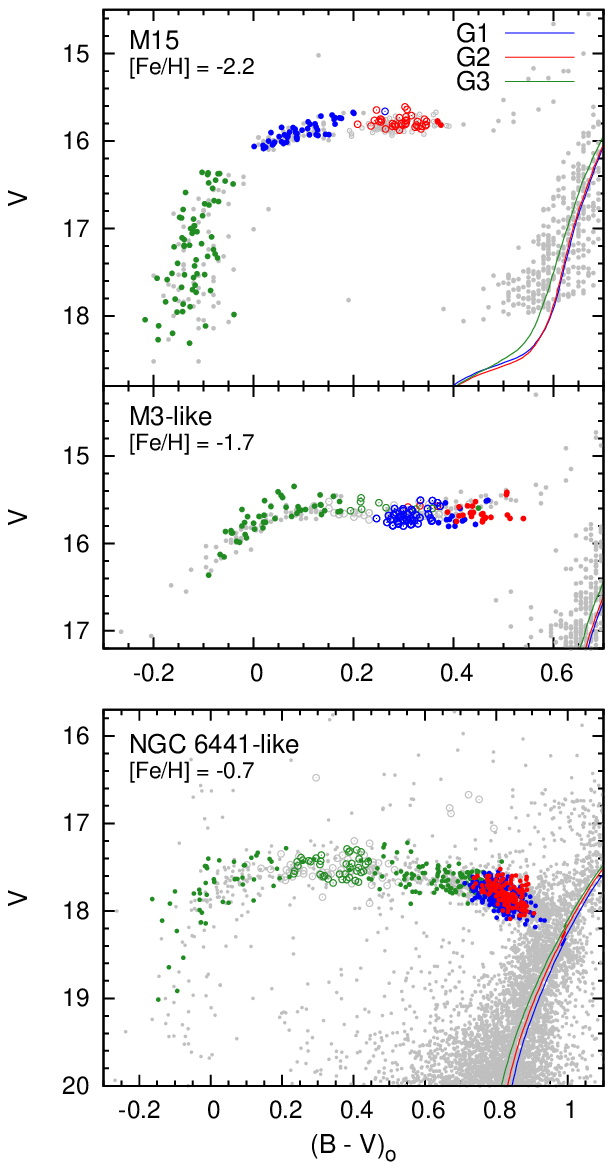}
 %\vspace*{174pt}
\caption{Comparison of our HB models for M15, M3-like, and NGC 6441-like with the observed CMDs (data from \citealt{Buo85,Buo94,Bin84,Pri03}). In our model for M15 (group II), the blue HB belongs to G1, while the RR Lyrae variables (open circles) are produced by helium and CNO enhanced G2. The blue tail hotter than normal blue HB is progeny of more helium-rich G3. When this model is shifted redward by increasing metallicity, the HB morphologies similar to M3 and NGC 6441 are obtained, and the instability strip becomes progressively populated by G1 and G3, respectively. Adopted distance modulus and reddening are ($m$~-~$M$)$_{\rm V}$ = 15.40, 15.24, \& 17.25 mag and $E$($B$~-~$V$) = 0.08, 0.03, \& 0.45 mag, for M15, M3, and NGC 6441, respectively.\label{fig:hb}}
\end{figure}

\begin{figure}
\centering
\vspace{2pt}
\includegraphics[width=0.47\textwidth]{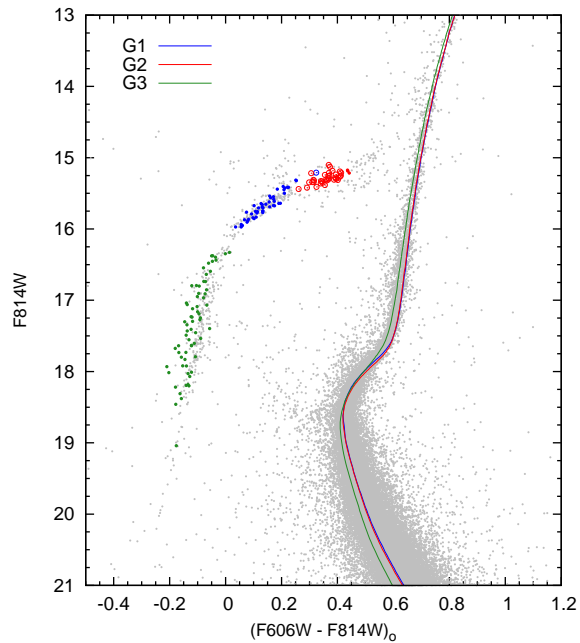}
%\includegraphics[angle=90,scale=.50]{fig3.eps}
%\epsscale{.80}
%\plotone{f1.eps}
\vspace{-5pt}
\caption{Comparison of our population model with the observed CMD for M15 \citep[data from][]{And08}. Adopted distance modulus and reddening are ($m$~-~$M$)$_{\rm F814W}$ = 15.35 mag and $E$(F606W~-~F814W) = 0.065 mag.\label{fig:acs}}
%\vspace{-10pt}
\end{figure}

Photometry of M15 shows, ignoring a few red HB stars, three distinct subgroups on the HB: RR Lyraes, blue HB, and the blue tail (see Fig.~\ref{fig:hb}, upper panel). Interestingly, other metal-poor group II GCs without the blue tail, such as NGC~5466 and NGC~4590, also show distinct gaps between the blue HB and RR Lyraes \citep{Cor99,Wal94}. We assume that these subgroups and gaps are originated from distinct subpopulations in these GCs. According to this picture, M15, for example, would contain three subpopulations, while NGC 5466 and 4590 are consisted with two subpopulations. Most of the colour spread on the HB is then due to the presence of multiple populations, rather than the mass dispersion \citep{Cat04}. Therefore, in our population modeling, the mass dispersion on the HB was assumed to be very small ($\sigma_{M}$ = 0.009 $M_{\rm \sun}$). Figure~\ref{fig:hb} (upper panel) and Figure~\ref{fig:acs} compare our models for M15 with the observations. Our models are based on the updated $Y^{\rm 2}$ isochrones and HB evolutionary tracks, including the cases of enhanced helium and CNO abundances. For the details of our model construction, readers are referred to \citet{Joo13}.

%%###############################################################

\begin{table*}

\setlength{\tabcolsep}{11pt}

\centering

\begin{minipage}{160mm}

%\begin{center}

\caption{Parameters from our best-fit simulation of M15}

%\begin{tabular}{@{}llrrrrlrlr@{}}

\begin{tabular}{@{}cccccccc@{}}

 \hline
 \hline

Population& {[Fe/H]\footnote{[$\alpha$/Fe] = 0.3 for G1 \& G2, 0.5 for G3.}}&  $\Delta$$Z_{\rm CNO}$ & $Y$ & Age & Mass Loss\footnote{Mean mass loss on the RGB for $\eta$ = 0.42.} &  $\left<M_{\rm HB}\right>$\footnote{Mean mass on the HB.} & Fraction \\
&&&&(Gyr)&($M_{\rm \sun}$)&($M_{\rm \sun}$)& \\

 \hline

G1&-2.2&0&0.230&12.5&0.140&0.686&0.36\\
G2&-2.2&0.00026&0.245~$\pm$~0.008&11.4~$\pm$~0.2&0.142&0.684&0.22\\
G3&-2.2&0&0.327~$\pm$~0.008&11.3~$\pm$~0.2&0.129&0.589&0.42\\

\hline

\vspace{-20pt}

\label{tbl:pm}

\end{tabular}

%\end{center}

\end{minipage}
\end{table*}

%%########################################################

In our modeling for M15, we start by placing the first generation stars (G1) on the HB (see Fig.~\ref{fig:hb}, upper panel). By adopting [Fe/H] = -2.2 \citep{Har96} and the \citet{Rei77} mass-loss parameter $\eta$ = 0.42\footnote{This value of $\eta$ is somewhat smaller than the value (0.53) suggested by \citet{Joo13}. This is because helium-enhanced subpopulations are included in their GC sample used in the $\eta$ calibration, while here we are adopting the value that would be obtained when they are excluded.}, we find G1 would be placed on the blue HB at the age of 12.5 Gyr. When the blue tail is excluded, the HB type of M15 becomes 0.46 \citep{Wal94}, which is too red to have enough period-shift from the evolution effect alone. This suggests that most, if not all, RR Lyraes in M15 are produced by the second generation stars (G2) with chemical compositions favorable to produce RR Lyraes with longer periods. Both theories and observations suggest G2 would be somewhat enhanced in both helium and CNO abundances, preferably in metal-poor GCs \citep{Ven09,Dec09,Alv12,Mar12,Gra12b}. While the enhanced helium abundance shifts the HB to blue, both CNO enhancement and younger age for G2 favor redder HB \citep[LDZ II;][]{Joo13}. Thus, when $\Delta$$Y$ is relatively small, the effects from CNO and younger age would overwhelm the helium effect, and the net effect will move the HB to red. We find that $\Delta$$Y$ = 0.015, $\Delta$$Z_{\rm CNO}$ = 0.00026 ($\Delta$[CNO/Fe] = 0.47 dex), and $\Delta$$t$ = 1.1 Gyr between G1 and G2 would best match the observations, both distribution of RR Lyraes on the CMD (see Fig.~\ref{fig:hb}, upper panel) and the period-shift (see below). Some evidence for the possible difference in CNO abundance between G1 and G2 in M15 is provided by \citet{Coh05}, who found a large spread in [N/Fe] ($\sim$~2~dex) among stars having identical [C/Fe] in M15. This suggests that CNO sum could be different if variation in [O/Fe] is not significant among them. It is also interesting to note that out of 6 RGB stars observed in M15, \citet{Sne97} found one ($\sim$~17~$\%$) shows significant enhancement ($\sim$~0.4~dex) in CNO abundance, roughly consistent with our ratio of G2 ($\sim$~20~$\%$). Certainly, spectroscopy with a large sample of stars, preferably at the lower RGB, is required to confirm this scenario. Finally, many previous works suggest that the blue tail (EBHB) would be produced by super-helium-rich subpopulation \citep{D'A04,Lee05,Gra12b,Joo13,Kun13a}, and therefore we assign more helium-rich third and later generations of stars (hereafter collectively G3) for the progenitor of EBHB. It is evident that some helium spread within G3 is required to reproduce the observed extension of the blue tail, which is mimicked here by increasing mass dispersion ($\sigma_{M}$ = 0.023 $M_{\rm \sun}$). Table~\ref{tbl:pm} lists our best fit input parameters for M15, where the population ratio is from distinct subgroups observed on the HB of \citet{Buo85}

It is interesting to see that when our HB model for M15 is shifted redward by increasing metallicity, the HB morphology similar to M3 (HB type = 0.08) is naturally obtained. This is illustrated by a more metal-rich model in Figure~\ref{fig:hb} (middle panel). For this model, identical values of ages, $\Delta$$Z_{\rm CNO}$, and mass-loss parameter $\eta$ adopted for M15 have been used, while the metal abundance varies from [Fe/H] = -2.2 to [Fe/H] = -1.7. The helium abundances for G1 and G2 have also been fixed, while that for G3 has been reduced to Y = 0.28, because otherwise our model would produce blue HB that is too blue compared to the observation of M3. Note that, because of the dispersion in helium abundance among G3, at the red end of the blue HB for M3, which is the critical regime for the \citet{Cat09b} test, the enhancement in Y would be only $\sim$~0.015 compared to the red HB. Because the overall HB morphology is shifted to red, the instability strip is now mostly occupied by G1, while the red HB is populated by both G1 and G2. Consequently, the gap, which was placed between the EBHB (G3) and blue HB (G1) in our M15 model, is likewise shifted into the instability strip, which agrees well with the observation for M3 \citep[see Fig. 14 of][]{Buo94}. Some evolved stars from G1, G2 and G3 are also placed within the instability strip, which would explain the presence of minority population of brighter RR Lyraes with longer periods observed in this GC \citep{Cac05}. Further fine tuning of input parameters, such as age, $Z_{\rm CNO}$, helium abundance, and population ratio for each subpopulation would be required to obtain a better match with the observation.

When the metallicity is increased further to [Fe/H] = - 0.7, our models yield the HB morphology that is analogous to NGC 6441 (HB type = - 0.77), the Oosterhoff III GC with longest $\left<P_{\rm ab}\right>$ (lower panel). Again, $\Delta$$t$ and $\Delta$$Z_{\rm CNO}$ among subpopulations, $\Delta$$Y$ between G1 and G2, and the mass-loss parameter $\eta$ are held identical to those adopted for M15. For this model, however, absolute ages have been increased by $\sim$~1 Gyr compared to M3-like model, and the mean value of the helium abundance for G3 has been adopted to be Y = 0.30. Note that the red HB is populated by all three subpopulations, while G3 is penetrating into the instability strip, producing RR Lyrae variables and some blue HB stars.

\begin{figure}
\centering
%\vspace{-15pt}
\includegraphics[width=0.43\textwidth]{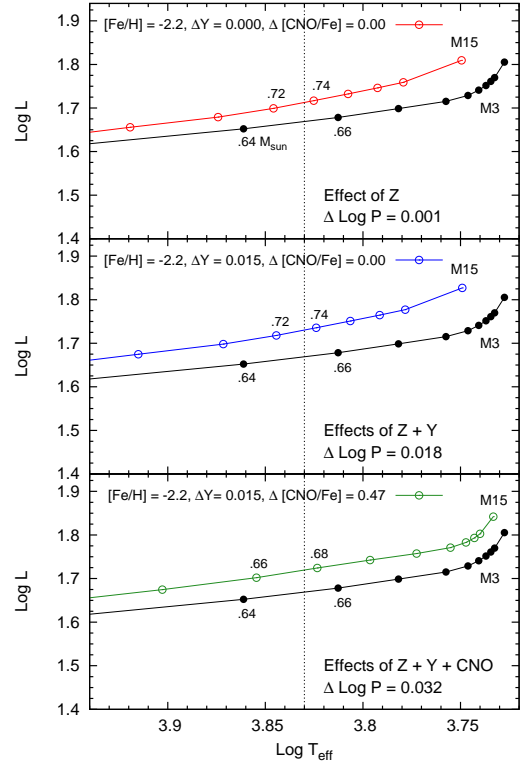}
\vspace{0pt}
\caption{Zero-age HB models explaining the origin of the period-shift between M15 and M3 at given temperature (log~$T_{\rm eff}$ = 3.83, vertical dashed line). While enhancement in helium increases luminosity (middle panel), enhancement in CNO reduces mass of HB stars at given temperature (lower panel). Both of these effects play a role in increasing the period-shift of RR Lyraes in M15 (see the text).\label{fig:zahb}}
\end{figure}

Figure~\ref{fig:zahb} explains, using ZAHB models, the origin of the Sandage period-shift effect between M15 (group II) and M3 (group I) in the new paradigm. In our model for M15, G2, which is enhanced in helium and CNO abundances, are in the instability strip. While enhancement in helium increases luminosity, enhancement in CNO reduces mass of HB stars at given temperature. Both of these effects play a role in increasing the period of RR Lyrae variables \citep{van73} in our model for M15. Consequently, in our synthetic HB models, where the full evolutionary tracks are employed in addition to ZAHB models, the period-shift between M15 and M3 is predicted to be $\Delta$log~$P$ = 0.040, where $P$ is in days. Adopting the temperature shift of the fundamental blue edge ($\Delta$log~$T_{\rm eff}$ = 0.02) as a function of metallicity \citep{San06}, the same models yield the difference in mean period of type ab RR Lyraes of $\Delta$$\left<P_{\rm ab}\right>$ = 0.087 day. These values agree well with the observed period-shift ($\Delta$log~$P$ = 0.044~$\pm$~0.01; LDZ I) and the difference in $\left<P_{\rm ab}\right>$ ($\Delta$$\left<P_{ab}\right>$ = 0.082~$\pm$~0.02 day; \citealt{Cle01}) between these GCs. Despite the uncertainty in the fundamental blue edge, the fraction of c type RR Lyraes ($f_{\rm c}$) is predicted to be 0.48 and 0.13 for M15 and M3, respectively, which should be compared with the observed values, 0.53 and 0.18 \citep{Cle01}. For our NGC 6441-like model, we obtain $\Delta$$\left<P_{ab}\right>$ = 0.18 day and $f_{\rm c}$ = 0.22, which is in reasonable agreement with the observation \citep[$\Delta$$\left<P_{ab}\right>$ = 0.20~$\pm$~0.02 day and $f_{\rm c}$ = 0.33;][] {Pri03}.

%%##############################################################################################################################

%%##############################################################################################################################

\begin{figure}
\centering
\vspace{-10pt}
\includegraphics[width=0.51\textwidth]{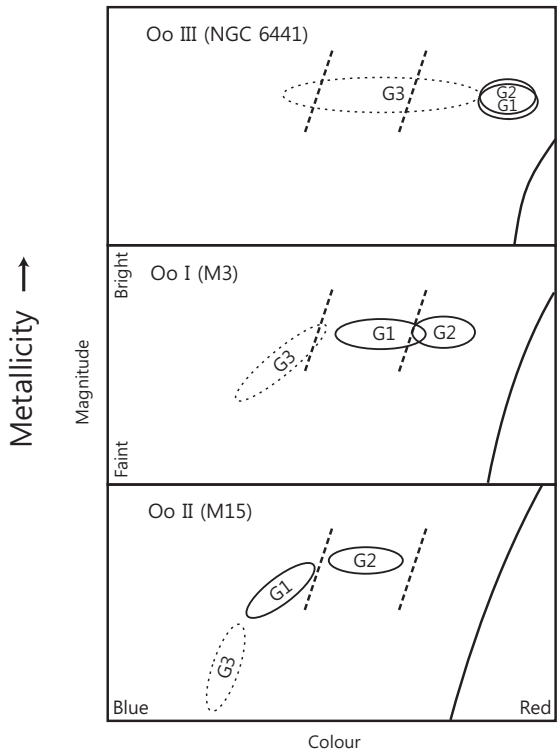}
\vspace{-3pt}
\caption{Schematic diagram illustrating the ``population shift'' within the instability strip (thick dashed lines) with increasing metallicity. For the most general cases, most of the RR Lyraes are produced by G1, G2 (helium \& CNO enhanced), and G3 (most helium-rich), respectively, for the Oosterhoff groups I, II, and III (see the text).\label{fig:oo}}
\end{figure}

\section{Discussion}

Figure~\ref{fig:oo} shows a schematic diagram that explains the main points from our models. As discussed above, RR Lyraes in the metal-poor group II cluster M15 are produced by helium and CNO enhanced G2 (lower panel). When the HB of this metal-poor model is shifted redward with increasing metallicity, the instability strip becomes mostly populated by G1 having Oosterhoff I characteristics (middle panel). When this redward shift continues as metallicity increases further, our models indicate that the instability strip would be more populated by G3, producing first the transition case between the groups I and III, having mildly helium enhanced (Y $\approx$~0.26) RR Lyraes with Oosterhoff-intermediate characteristics, such as NGC 1851, as suggested by \citet{Kun13a, Kun13b}. Then, in the most metal-rich regime, if G3 are present, GCs with RR Lyraes having more enhanced helium abundance (Y $\approx$~0.30) and longest periods like NGC 6441 (group III) would be produced (upper panel). Note that the placements of G1 and G2 in this schematic diagram is valid only for GCs where $\Delta$$t$ between G1 and G2 is similar to that in the case of M15. If $\Delta$$t$ is much smaller than $\sim$~1 Gyr as adopted in our models, for example, the RR Lyraes in group I GCs would be more dominated by G2, while the red HB becomes more populated by G1. It appears unlikely, however, that these variations are common, because otherwise most group I GCs would have Oosterhoff-intermediate characteristics in period-shift and $\left<P_{\rm ab}\right>$.

Some GCs like NGC 6397 show very small spread in the colour on the HB, as well as in the Na-O plane \citep{Car09}, which suggest that these GCs are probably consisted with only G1. Therefore, there are cases/systems where the suggested population shift cannot be a viable explanation for the Oosterhoff dichotomy. For these GCs, evolution away from ZAHB (LDZ I) and some hysteresis mechanism \citep{van73} are probably at works for the difference in $\left<P_{\rm ab}\right>$ between the groups I and II.

The Na-O anticorrelations observed in some HB stars in GCs can provide an important test on our placements of G1, G2, and G3 on the HB . For example, for M5, when the division of G1 and G2 is made properly at [Na/Fe] = 0.1 as suggested by \citet{Car09}, Fig. 9 of \citet{Gra13} shows that $\sim$~65~$\%$ of stars in the red HB have G2 characteristic, while $\sim$~35~$\%$ are in G1 regime. This is in agreement with the ratio of G2/G1 ($\sim$~2; see Fig. 1) in the red HB of our model for M3-like GCs. For more metal-rich GCs, our models predict that the red HB is roughly equally populated by G1 and G2, which agrees well with the Na-O observations for NGC 1851 and NGC 2808 \citep{Gra12b,Mar14}. For the blue HB stars, these observations confirm that they belong to Na-rich and He-rich G3. It is interesting to note that one RR Lyrae variable in NGC 1851 was observed to be Na-rich like blue HB stars (G3), which is consistent with our suggestion that NGC 1851 is a transition case between the Oosterhoff I and III. For the group II GCs, the Na - O observation of HB stars is available only for M22 \citep{Gra14}, which shows that the blue HB stars right next to RR Lyraes have G1 characteristic. While this is in qualitative agreement with our model for M15, the interpretation is more complicated as this GC was also affected by supernovae enrichment \citep[and references therein]{Joo13}. Certainly, spectroscopic observations for a large sample of RR Lyrae and horizontal-branch stars in GCs representing two Oosterhoff groups, such as M15 and M3, are urgently required to confirm our models.

%%##############################################################################################################################

\section*{Acknowledgments}

We thank the referee for a number of helpful suggestions. We also thank Robert Zinn, Pierre Demarque, Suk-Jin Yoon, Chul Chung, Sang-Il Han, and Dong-Goo Roh for helpful discussions. Support for this work was provided by the National Research Foundation of Korea to the Center for Galaxy Evolution Research.\\

%%##########################################################################################################

%%###########################################################################################################

\bsp

\label{lastpage}

\end{document}